\documentclass[aps,prb,reprint,superscriptaddress,hidelinks,nobalancelastpage,longbibliography]{revtex4-1}% for checking your page length
\usepackage{amsmath,braket}
\usepackage{graphicx}
\usepackage{color}
\usepackage{hyperref}
\usepackage[separate-uncertainty=true,multi-part-units=single]{siunitx}
\usepackage{changes}

\begin{document}
\title{Phonon-assisted dark exciton preparation in a quantum dot}

\author{S.~L\"uker}
\affiliation{Institut f\"ur Festk\"orpertheorie, Universit\"at M\"unster,
Wilhelm-Klemm-Str.~10, 48149 M\"unster, Germany}

\author{T.~Kuhn}
\affiliation{Institut f\"ur Festk\"orpertheorie, Universit\"at M\"unster,
Wilhelm-Klemm-Str.~10, 48149 M\"unster, Germany}

\author{D.~E.~Reiter}
\affiliation{Institut f\"ur Festk\"orpertheorie, Universit\"at M\"unster,
Wilhelm-Klemm-Str.~10, 48149 M\"unster, Germany}

\date{\today}

\begin{abstract}
In semiconductor quantum dots the coupling to the environment, i.e. to phonons, plays a crucial role for optical state preparation. We analyze the phonon impact on two methods for a direct optical preparation of the dark exciton, which is enabled by a tilted magnetic field: the excitation with a chirped laser pulse and the excitation with a detuned pulse. Our study reveals that for both methods phonons either do not impede the proposed mechanism or are even useful by widening the parameter range where a dark state preparation is possible due to phonon-assisted dark exciton preparation. In view of the positive impact of phonons on the optical preparation the use of dark excitons in quantum dots becomes even more attractive.
\end{abstract}

%\pacs{78.67.Hc, 78.47.D-, 42.50.Md}

%\keywords{quantum dots; dark excitons; state preparation; adiabatic rapid passage}

\maketitle

\section{Introduction}
%%%%%%%%%%%%%%%%   Intro   %%%%%%%%%%%%%%%%%%%%%%%%%%%%%%%%%
The optical control of a quantum state in a semiconductor quantum dot (QD) is of great interest for applications in quantum information. While phonons often limit the preparation fidelity, e.g., in the case of resonant preparation by Rabi rotations \cite{ramsay2010dam,reiter2014the}, recently, phonons have been actively used in optical control schemes of QD states \cite{reiter2012pho,glassl2013pro,bounouar2015pho,ardelt2014dis,quilter2015pho}. In these protocols, the emission of phonons was exploited to prepare the exciton \cite{reiter2012pho,quilter2015pho,manson2016pol} or biexciton state of the QD \cite{glassl2013pro,bounouar2015pho,ardelt2014dis,barth2016fas} or to depopulate the QD \cite{liu2016ult}. However, all of these excitation protocols act on the optically active or \textit{bright} exciton, while the optically inactive or \textit{dark} exciton does not come into play. Due to the lack of optical activity dark excitons possess significantly longer lifetimes than bright excitons \cite{stevenson2004tim}, which offers potential for applications in quantum memory, but also makes their preparation more challenging. Due to these prospects several mechanisms for the controlled preparation of the dark exciton have been recently studied either by relaxations from higher excited states \cite{korkusinski2013ato,poem2010acc,smolenski2015mec,schmidtgall2015all} or valence band mixing \cite{schwartz2015det}. We recently proposed a scheme to direct optical address the dark exciton by using a combination of a single chirped laser pulse and a tilted magnetic field and we also showed that at sufficiently low temperatures the process is essentially unaffected by phonons \cite{luker2015dir}. The question remains, whether also for dark excitons phonon-assisted state preparation schemes - in analogy to preparation schemes of the bright exciton-- can be employed. 

In this paper, we present two different optical excitation schemes by which the dark exciton can be prepared in a tilted magnetic field making active use of phonons: (i) an excitation with a chirped laser pulse and (ii) an excitation with a detuned laser pulse having constant frequency. For both excitation scenarios we show that phonon assisted state preparation is possible and discuss the relevant excitation conditions. By comparing our results to the phonon-free case, we show that phonons are not hindering the dark state preparation at low temperatures, but rather widen the parameter range where a dark state preparation is possible. To analyze the results, we study the dynamics in the dressed state basis, giving fundamental insight into the phonon influence on the dynamics of an optically driven few-level system.

%%%%%%%%%%%%%%%   Theory   %%%%%%%%%%%%%%%%%%%%%%%%%%%%%%%%%%

\section{Theoretical model}
\subsection{Quantum dot model}
\begin{figure}[t]{}
	\centering
	\includegraphics[width=1.0\columnwidth]{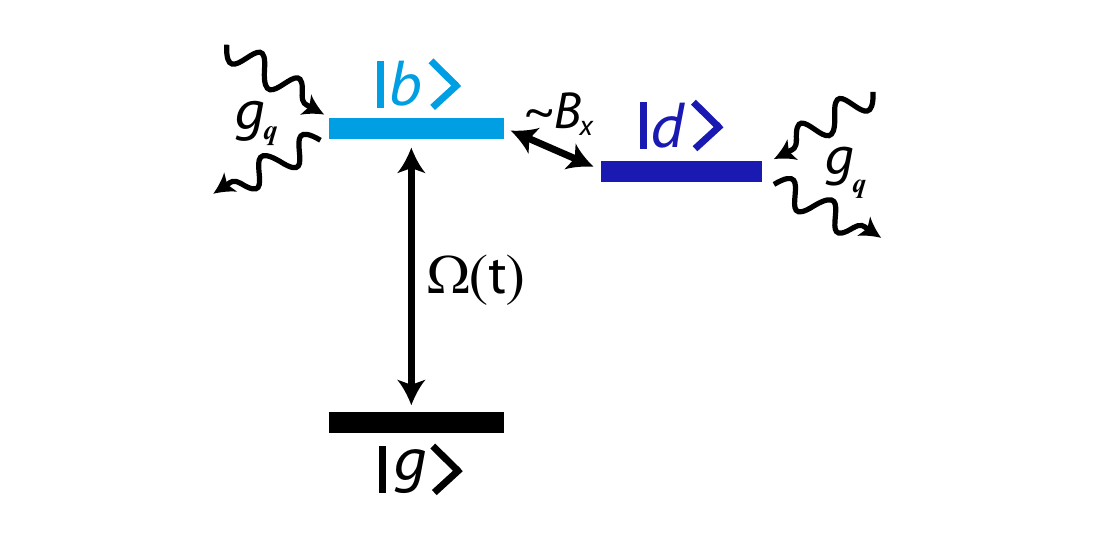}
      \caption{Sketch of the three-level system including the interactions between the states and coupling to phonons.}
	\label{fig1}
\end{figure}

Considering s-shell heavy-hole excitons,  the bright excitons consist of an electron with spin $S_z^e=\pm1/2$ and a hole with spin $S_z^h=\mp3/2$ being anti-parallel, which results in a total spin of $J_z^b=\pm1$.  Analogously, the dark exciton consists of a parallel spin configuration with total spin $J_z^d=\pm2$. According to the dipole selection rules a photon can only transfer an angular momentum of $\pm 1$ making the states with $J_z^b=\pm1$ optically active, while the states with $J_z^d=\pm2$ are not optically accessible. The most general model of the ground state excitons in a QD thus consists of six states, i.e., the ground state $|g\rangle$ without an exciton in the dot, two bright single excitons $|b^+\rangle$ and $|b^-\rangle$, two dark single excitons $|d^+\rangle$ and $|d^-\rangle$, and the biexciton $|B\rangle$. We restrict ourselves on the excitation with negative circularly polarized laser pulses. When the system is initially in the ground state, these pulses can only excite the negatively polarized bright exciton $|b^-\rangle$. Furthermore, we apply a magnetic field with a z-component, which shifts the positively polarized states far out of resonance from the negatively polarized states, such that the coupling of $|b^-\rangle$ ($|d^-\rangle$) with $|b^+\rangle$ ($|d^+\rangle$), which is typically present in many QDs due to the long-range exchange interaction, is strongly suppressed. Therefore, the system reduces to a three-level system formed by the ground state $|g\rangle$, one bright exciton $|b\rangle$ and one dark exciton $|d\rangle$, where we suppress the polarization index for brevity. We have checked that for the parameters used here the restriction to three levels is indeed very well satisfied \cite{luker2015dir}. The energetic structure of the system is then described by
\begin{equation}
H_c = \hbar\omega_b |b\rangle\langle b| + \left(\hbar\omega_b - \delta_0 + g_{e,z}\mu_BB_z\right)|d\rangle\langle d|,
\end{equation}
where we have chosen the energy of the ground state to zero, $\hbar\omega_b$ is the energy of the bright exciton. The short-range exchange splitting $\delta_0 = 0.25\,{\rm{meV}}$ leads to a shift of the energy of the dark exciton with respect to the bright exciton. A magnetic field with z-component $B_z$ enlarges the energy splitting and therefore prevents a backflip into the bright exciton, $\mu_B$ is the Bohr magneton. This additional energy shift is solely induced by the electron via its out-of-plane g-factor $g_{e,z} = -0.8$. The energy structure of the three-level system is sketched in Fig.~\ref{fig1}.

The coupling of the QD to the light field is modeled in the usual rotating wave and dipole approximation. The corresponding Hamiltonian reads
\begin{equation}
H_{c-l} = \frac{\hbar}{2}\Omega(t)e^{-i\varphi(t)}|b\rangle\langle g| + \frac{\hbar}{2}\Omega^*(t)e^{i\varphi(t)}|g\rangle\langle b|,
\end{equation}
with $\Omega(t) = 2\mathbf{E}(t)\cdot\mathbf{M}/\hbar$, where $\mathbf{M}$ is the dipole matrix element and $\mathbf{E}(t)$ describes the electric field of the laser pulse with instantaneous frequency $\omega_L = \frac{d\varphi}{d t}$. Considering the excitation with Gaussian laser pulses, the envelope of the pulse reads $\Omega(t) = \frac{\Theta}{\sqrt{2\pi}\tau_0}\exp\left(-t^2/2\tau_0^2\right)$, where $\Theta$ denotes the pulse area and $\tau_0$ is the pulse duration.
The x-component $B_x$ of the magnetic field provides a coupling between the bright and dark exciton via spin flip of the electron, yielding
\begin{equation}
H_{mag} = \frac{g_{e,x}\mu_B B_x}{2}\left(|b\rangle\langle d| + |d\rangle\langle b|\right),
\end{equation}
where $g_{e,x} = -0.65$ denotes the in-plane g-factor of the electron. The ratio between the in-plane and out-of-plane magnetic g-factor is a crucial parameter for the preparation of the dark exciton. On the one hand, the in-plane magnetic field has to be strong enough to enable a transition to the dark exciton. On the other hand, the out-of-plane magnetic field should be strong enough to ensure a clear distinction between bright and dark exciton.

\subsection{Coupling to phonons}
The coupling to phonons of the surrounding bulk material is the main source of decoherence in self-assembled QDs. Previous studies have shown that longitudinal acoustic (LA) phonons coupled via the deformation potential coupling mechanism dominate the dephasing in optically driven, self-assembled QDs \cite{woggon2004non,krummheuer2002the,ramsay2010dam,reiter2014the}, while the coupling via the piezoelectric mechanism, which would allow a coupling to transverse acoustic phonons, is typically rather weak \cite{krummheuer2002the}. Thus, we restrict ourselves on the coupling to LA phonons. The Hamiltonian of the free phonons reads
\begin{equation}
H_{ph} = \hbar\sum_{\mathbf{q}}\omega_{\mathbf{q}}b^\dag_{\mathbf{q}}b_{\mathbf{q}},
\end{equation}
where $b_{\mathbf{q}}$ ($b^\dag_{\mathbf{q}}$) is the annihilation (creation) operator for a phonon with wave vector $\mathbf{q}$ and frequency $\omega_{\mathbf{q}} = c_{LA}q$, with $c_{LA}$ being the sound velocity. All calculations are performed for a phonon-bath temperature of $T = 1\,{\rm{K}}$. 
The carrier-phonon coupling is modeled using the deformation potential mechanism via a pure-dephasing Hamiltonian
\begin{equation} H_{c-ph} = \hbar\left(|b\rangle\langle b| + |d\rangle\langle d|\right)\sum_{\mathbf{q}}\left(g_{\mathbf{q}}b_{\mathbf{q}} + g_{\mathbf{q}}^*b_{\mathbf{q}}^\dag\right),
 \end{equation}
where the $g_{\mathbf{q}}$ denote the deformation potential coupling constants using standard GaAs parameters \cite{krugel2006bac}. Because the coupling to the phonons is not affected by the spin of the carriers, the coupling constants are the same for bright and dark exciton.
The carrier-phonon coupling leads to an energy shift of the bright and dark exciton due to the formation of a polaron. The polaron energy reads $\hbar\omega_{pol} = -\hbar\sum_{\mathbf{q}}|g_{\mathbf{q}}|^2/\omega_{\mathbf{q}} \approx -0.17\,{\rm{meV}}$.
We use the density matrix formalism to calculate the dynamics. Due to the many-body character of the carrier-phonon interaction, there is no closed set of equations of motion, but an infinite hierarchy of higher order phonon-assisted density matrices. To truncate this hierarchy we use a fourth-order correlation expansion \cite{krugel2006bac,luker2012inf}, which provides a treatment of the carrier-phonon coupling including non-Markovian effects. Our method has shown excellent agreement with experiments \cite{kaldewey2016dem,kaldewey2017coh} and also with a numerically exact path-integral method for a wide range of parameters \cite{glassl2011lon}.

\subsection{Dressed states}
An intuitive picture to study the dynamics during the excitation is provided by the dressed states, i.e., the instantaneous eigenstates of the system in a frame rotating with the laser frequency $\omega_L$. The diagonalization is performed on the reduced Hamiltonian $\widetilde{H} = H_c + H_{c-l} + H_{mag}$ resulting in the dressed states $|i\rangle$ $(i=1,2,3)$. To account for the electron-phonon interaction, at this stage we only take into account the polaron shift reducing the energies of the bright and dark exciton by $\tilde{E}_{b/d}=E_{b/d}- \hbar\omega_{pol}$. The electron-phonon interaction also leads to transitions between the eigenenergies, which will be discussed below. The time evolution in the  dressed states is obtained by considering the instantaneous eigenenergies at each time $t$. 

An \emph{adiabatic} evolution occurs, when the system remains on the same eigenenergy branch for all times, i.e., such that no transitions between the eigenstates occur. A criterion for adiabatic evolution is provided by the Landau-Zener formalism: In a nutshell, the system will remain in the prepared eigenstate, if the variation of the energy is sufficiently slow and the energy difference to other eigenenergies is sufficiently large \cite{landau1932ath,zener1932non,tannor2007int}. For the two-level system, the excitation with a chirped laser pulse then leads to the \emph{adiabatic rapid passage} (ARP) effect, resulting in a stable population inversion. The ARP effect in QDs has been recently demonstrated \cite{wu2011pop,simon2011rob,mathew2014sub,kaldewey2017dem}. 

If two branches of eigenenergies come very close to each other or the evolution along the branches is very fast, a \emph{diabatic} transition occurs, where the system jumps from one branch to another. The diabatic transition probability can be calculated in the Landau-Zener formalism \cite{tannor2007int}. By performing the numerical calculations in the original (exciton) basis, both adiabatic and diabatic processes are fully included.

In the dressed states, the electron-phonon interaction can lead to transitions between the different eigenstates, where the transition from an upper to a lower state occurs with the emission of a phonon, while the reverse process is accompanied by the absorption of a phonon. At low temperatures, as considered here, phonon absorption processes are negligible, because in the required spectral range there are no phonons to be absorbed. This picture was successfully used to explain the asymmetry of phonon decoherence for the ARP effect with respect to the sign of the chirp \cite{luker2012inf,kaldewey2016dem,mathew2014sub,eastham2013lin} and to explain phonon-assisted bright exciton preparation \cite{quilter2015pho,barth2016fas}. For higher temperatures, phonon absorption and emission lead for sufficiently long pulses to an almost equal occupation of the dressed states, which hinders a high-fidelity state preparation using ARP \cite{luker2012inf,eastham2013lin} or phonon-assisted processes \cite{reiter2012pho}.

%%%%%%%%%%%%%%%   Results: chirped  %%%%%%%%%%%%%%%%%%%%%%%%%%%%%%%%%%
\section{Results}
\subsection{Excitation with chirped pulses}

In this section we discuss the dark exciton preparation using a single chirped laser pulse. In a chirped pulse the laser frequency $\omega_L$ changes in time according to $\omega_L(t) = \omega_b + \omega_{pol} + \delta\omega + at$, where $\delta\omega$ denotes the detuning at $t = 0$ with respect to the polaron shifted bright exciton, the chirp rate $a=\frac{\alpha}{\alpha^2+\tau_0^4}$ is determined by the chirp coefficient $\alpha$ of the chirp filter used to create the chirped pulse from a transform limited Gaussian one. The envelope of the chirped pulse reads $\Omega(t) = \frac{\Theta}{\sqrt{2\pi\tau_0\tau}}\exp\left(-t^2/2\tau^2\right)$, where the pulse duration is increased by the chirp to $\tau = \sqrt{\alpha^2/\tau_0^2+\tau_0^2}$ \cite{luker2012inf}. We choose $\tau_0=2$~ps and a small detuning of $\hbar \delta \omega=0.5$~meV, which is helpful to analyze the different steps of the dynamics in detail. Nevertheless, the presented preparation mechanism also works without this additional detuning. For the magnetic field we take $B_x = 2.4\,{\rm{T}}$ and $B_z = 6.6\,{\rm{T}}$, which corresponds to a total field strength of $|\mathbf{B}|=7\,{\rm{T}}$ and a tilt angle of $20^\circ$.

\begin{figure}[t]{}
	\centering
	\includegraphics[width=1\columnwidth]{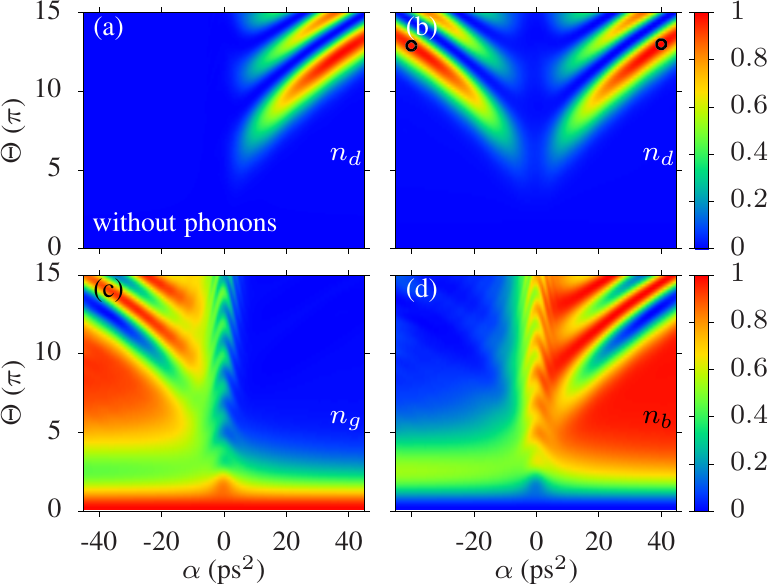}
      \caption{Final occupation of the states after excitation with a laser pulse with pulse area $\Theta$ and chirp coefficient $\alpha$. (a) Dark exciton occupation $n_d$ for the system without phonons. Occupation of (b) the dark exciton $n_d$, (c) the ground state $n_g$ and (d) the bright exciton $n_b$ including phonons. The small black circles refer to the parameters used in Fig.~\ref{fig3}.}
	\label{fig2}
\end{figure}

We start the discussion of the dark exciton preparation without the electron-phonon interaction. Figure~\ref{fig2}(a) shows the final occupation of the dark exciton $n_d$ for the phonon-free system as a function of the pulse area $\Theta$ and the chirp coefficient $\alpha$ of the laser pulse. We find a strong asymmetry of the dark exciton population with respect to the sign of the chirp: for negative chirps the dark exciton remains unoccupied and for positive chirps a stripe pattern appears, clearly showing a high dark exciton occupation for certain excitation parameters. This mechanism of direct dark state preparation has been explained in detail in Ref.~\onlinecite{luker2015dir}.

Let us now include the coupling to phonons for the calculation of the dark exciton occupation in Fig.~\ref{fig2}(b). Here, we find that for both, positive and negative chirp, the dark exciton becomes occupied and a stripe pattern symmetric with respect to the sign of the chirp comes up. This is in striking contrast to the influence of phonons on the bright exciton preparation, where an asymmetry at low temperatures is induced due to the lack of phonon absorption processes \cite{glassl2013pro,quilter2015pho,wei2014det}. However, there is a difference in the occupation the ground state $n_g$ and the bright exciton $n_b$, as shown in Fig.~\ref{fig2}(c) and (d). For negative chirp the occupation of the ground state shows a stripe pattern, which is complementary to the population of the dark exciton, i.e. here an oscillation as a function of the pulse area between the dark exciton and the ground state takes place, while the bright state occupation stays low for most pulse areas. For positive chirp the ground state is completely depleted for sufficiently high pulse areas and the bright exciton occupation exhibits a stripe pattern, showing that in this case an oscillation between dark and bright state occurs.

\begin{figure}[ht]{}
	\centering
	\includegraphics[width=1.0\columnwidth]{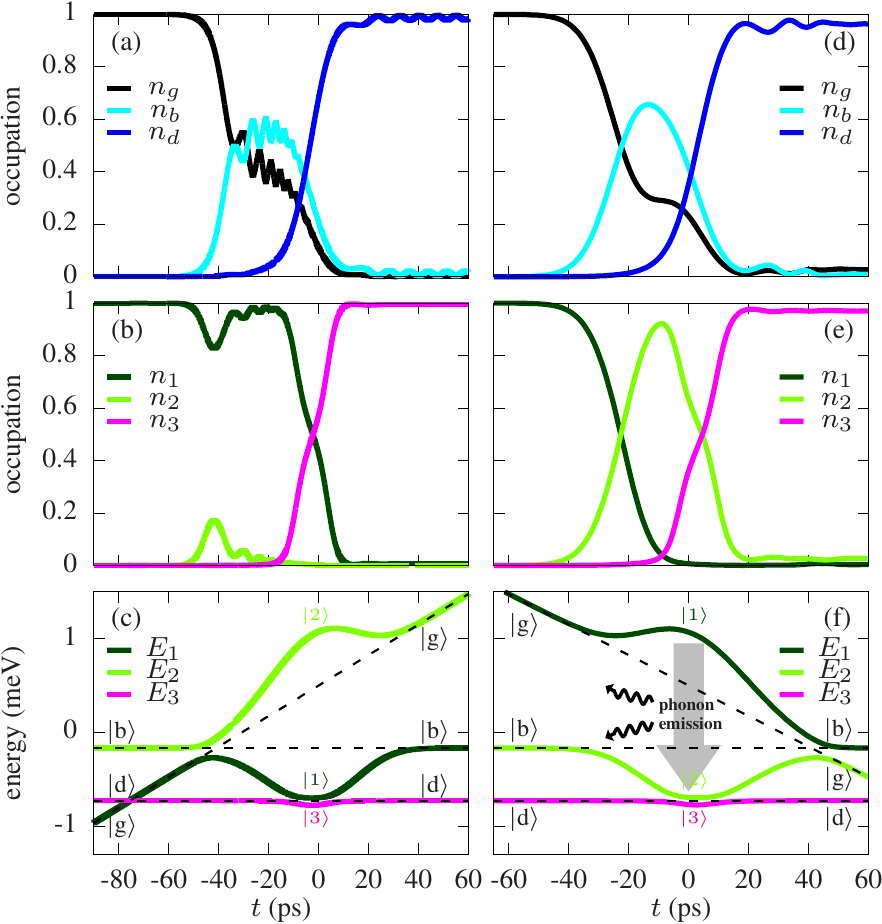}
      \caption{(a) Temporal evolution of the occupation of the ground state $n_g$, bright exciton $n_b$, and dark exciton $n_d$ during the excitation with a laser pulse having an area $\Theta = 13\pi$ and a positive chirp coefficient $\alpha = 40\,{\rm{ps}}^2$. (b) Temporal evolution of the occupation of the dressed states $n_i$ ($i=1,2,3$). (c) Instantaneous eigenenergies; the dashed lines indicate the eigenenergies in absence of coupling. (d), (e), and (f) show the same for negative chirp with $\alpha = -40\,{\rm{ps}}^2$.}
	\label{fig3}
\end{figure}

To understand the dark exciton preparation, we consider the temporal evolution of the system during the excitation. We start by discussing the dynamics for positive chirp, where the electron-phonon interaction does not play a crucial role as can be seen from comparing Fig.~\ref{fig2}(a) and (b). Figure~\ref{fig3}(a) shows the dynamics during the exciton with a $13\pi$-pulse with $\alpha = 40\,{\rm{ps}}^2$, which corresponds to the first maximum in Fig.~\ref{fig2}(b) as indicated by the black circle. Before the laser pulse, the system is in the ground state ($n_g = 1$). During the pulse $n_g$ drops down, while a transient occupation of the bright exciton occurs between $-40\,{\rm{ps}}$ and $0\,{\rm{ps}}$. Around $t = 0\,{\rm{ps}}$ a population transfer to the dark exciton takes place. Finally, the dark exciton is almost completely occupied, while the weak oscillations of $n_d$ are a result of a small admixture of the bright exciton. The corresponding dynamics of the dressed states is shown in Fig.~\ref{fig3}(b) and the eigenenergies are depicted in Fig.~\ref{fig3}(c). In absence of any coupling, i.e., before and after the pulse, each dressed state is identical to one bare state of the system. Before the laser pulse, the dressed states can be identified with the bare states, i.e., $|1\rangle = |g\rangle$, $|2\rangle = |b\rangle$, and $|3\rangle = |d\rangle$. When we forget about the dark state for a moment, we can identify the ARP effect in a two-level system: By the excitation with chirped laser pulses the actual detuning of the laser frequency with respect to the transition frequency changes with time and correspondingly changes the instantaneous splitting between $|g\rangle$ and $|b\rangle$. Due to the coupling an anti-crossing occurs and as a result, the character of the two dressed states, which are associated with the optically active states, changes after the pulse, i.e., $|1\rangle = |b\rangle$ and $|2\rangle = |g\rangle$. The adiabatic evolution along the dressed states then leads to the stable population inversion. Now, let's take state $|3\rangle$ back on board: Around $t \approx -75\,{\rm{ps}}$ a crossing between the branches of $E_1$ and $E_3$ occurs. Because the ground state is not directly coupled to the dark exciton, the evolution just follows the branches. More interesting is the narrow anti-crossing between $E_1$ and $E_3$, which occurs around $t = 0\,{\rm{ps}}$. The small energy difference allows for an efficient diabatic population transfer between the corresponding states, allowing the system to jump from $|1\rangle$ to $|3\rangle$. This is reflected in the temporal evolution of the dressed state occupation. While the system remains mostly in state $|1\rangle$ at the beginning, around $t=0$ the occupation swaps from $|1\rangle$ to $|3\rangle$, resulting in an occupation of the dark exciton. We further note that around $t=-40\,{\rm{ps}}$ approximately $20\%$ of the occupation goes into state $|2\rangle$, where partly a diabatic transition takes place. Since $|2\rangle$ corresponds to the highest eigenenergy, the population can be transferred to a lower branch by emission of a phonon. Thus the population returns successively back to $|1\rangle$. This is the only time, when phonons do play a role for the state preparation for positive chirp, at all other times, the evolution is not affected by the electron-phonon interaction. This is because the evolution takes place on the lowest eigenstates and no absorption of phonons is possible. 

One may wonder, if a phonon-assisted transition to $|3\rangle$ takes place, because for most of the time the eigenenergy $E_3$ is even the lowest one, such that in principle the relaxation to $|3\rangle$ should be even more favorable. The suppression of transitions to the lowest dressed state can be explained by the composition of the dressed states. An intuitive picture for the deviation from the bare states is provided by three dashed lines in Fig.~\ref{fig3}(c), which indicate the energy of $|g\rangle$, $|b\rangle$, and $|d\rangle$ in the limit of a vanishing laser amplitude. When the laser pulse sets in, the state $|1\rangle$ is mostly composed by the ground state, while it contains a small contribution of the bright exciton. The composition of the state $|2\rangle$ is complementary to $|1\rangle$, since it is dominated by the bright exciton and has a small admixture of the ground state. In contrast, the state $|3\rangle$ is almost identically to the dark exciton, which is indicated by the perfect alignment of $E_3$ and the lowest dashed line until $t \approx -20\,{\rm{ps}}$. Because the phonon coupling is of pure dephasing type, no real transitions between the bare states are allowed. Thus, a phonon-assisted transition has to be mediated by an additional coupling, i.e., by the laser pulse or by the magnetic field. Since the ground state is not directly coupled to the dark exciton, a transition from the ground-state dominated eigenstate $|1\rangle$ to $|3\rangle$ is strongly suppressed. On the other hand, the coupling between the bright and dark exciton should allow transitions between $|2\rangle$ and $|3\rangle$, once the system has left $|1\rangle$, i.e., around $t\approx -40\,{\rm{ps}}$. The fact that the phonon-assisted relaxation from $|2\rangle$ occurs exclusively to $|1\rangle$, but not to $|3\rangle$, originates from the identical coupling strengths of $|b\rangle$ and $|d\rangle$ to the phonons. Thus, the phonon coupling between the associated dressed states compensates each other, resulting in a suppression of phonon-assisted transitions between these states.

%%negative Chirp
Now we come to the dynamical behaviour for negative chirps shown in  Fig.~\ref{fig3}(d)-(f), where we consider an excitation with a $13\pi$-pulse with chirp coefficient $\alpha = -40\,{\rm{ps}}^2$, as indicated by a black circle in Fig.~\ref{fig2}(b). The dynamics of the bare states evolves similar to the positively chirped excitation. The occupation of the ground state drops down in favour of the bright exciton, which reaches a maximal occupation of $n_b \approx 0.7$ around $t = -10\,{\rm{ps}}$. Afterwards, the dark exciton becomes occupied. Even though the dynamics of the bare states looks similar to the positively chirped excitation, there is a big difference in the dynamics of the dressed state occupation (Fig.~\ref{fig3}(e)), most notably that the occupation of $|2\rangle$ rises up to nearly one during the pulse. To understand the difference, we look at the eigenenergies displayed in Fig.~\ref{fig3}(f). Starting again in $|1\rangle$, we are now in the uppermost eigenenergy due to the reversed chirp. From this state, phonon emission can take place to state $|2\rangle$. This process already is almost completed at $t=-10\,{\rm{ps}}$, such that when the anti-crossing between $E_2$ and $E_3$ around $t = 0\,{\rm{ps}}$ takes place, the system is almost completely in state $|2\rangle$. At this anti-crossing again diabatic transitions take place, which leads to a swap of occupation from $|2\rangle$ to $|3\rangle$, resulting in the occupation of the dark exciton.

We can thus summarize, that in comparison to the positively chirped excitation, which is basically a one-step process involving two dressed states, the dark exciton preparation using negatively chirped laser pulses has two components: first a phonon-assisted relaxation takes place, which is followed by a diabatic transition into the dark exciton state. The dressed state picture also explains the different behaviour in the ground and bright state occupation: While for positive chirp the system ends up in state $|1\rangle$ and $|3\rangle$, which correspond to $|b\rangle$ and $|d\rangle$, for negative chirp the oscillations take place between states $|2\rangle$ and $|3\rangle$, resulting in the bare states $|g\rangle$ and $|d\rangle$. Accordingly we find the difference in occupations in Fig.~\ref{fig2}(c) and (d).

\begin{figure}[t]{}
	\centering
	\includegraphics[width=1.0\columnwidth]{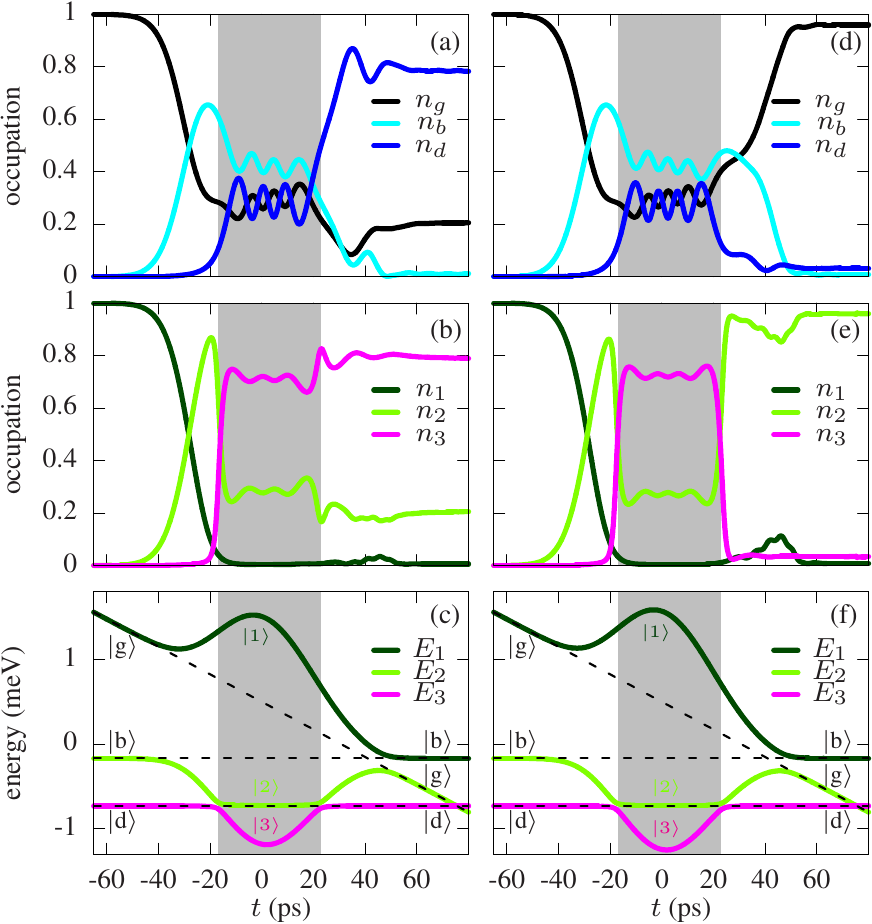}
      \caption{Same as Fig.~\ref{fig3}, but for $\alpha = -40\,{\rm{ps}}^2$ and $\Theta = 20\pi$ (left column) or $\Theta = 21\pi$ (right column). The gray background highlights the region between the anti-crossings of $|2\rangle$ and $|3\rangle$.}
	\label{fig4}
\end{figure}

When the pulse area is further increased, we find an oscillatory behaviour of the dark state occupation in Fig.~\ref{fig2}, which is due to oscillations induced by the diabatic transitions of the anti-crossing between $|2\rangle$ and $|3\rangle$. It is thus interesting to study the dynamics at higher pulse areas. Accordingly we show in Fig.~\ref{fig4} the dynamics for $\Theta = 20\pi$ and $\alpha = -40\,{\rm{ps}}^2$ (left column) and $\Theta = 21\pi$ and $\alpha = -40\,{\rm{ps}}^2$ (right column), corresponding to the 4th maximum and to the 4th minimum, respectively. For the dynamics of the bare states (Fig.~\ref{fig4}(a) and (d)) we can discriminate three different phases and explain them using the dressed state occupation (Fig.~\ref{fig4}(b) and (e)) and eigenenergies (Fig.~\ref{fig4}(c) and (f)):

In the first phase a decrease of the ground state occupation $n_g$ in favor of the bright state occupation $n_b$ takes place. Here, the phonon-assisted relaxation leads to a relaxation of the system from $|1\rangle$ into $|2\rangle$, in agreement with the change of occupation from $n_g$ to $n_b$. In the second phase between $t \approx -18\,{\rm{ps}}$ and $t \approx +22\,{\rm{ps}}$ we see oscillations between all three bare states, i.e., the occupations of the ground and bright state oscillate against the occupation of the dark state. 

The second phase is indicated by the gray shaded background. At the beginning of this phase around $t \approx -18\,{\rm{ps}}$ we see a diabatic transition from $|2\rangle$ to $|3\rangle$, which leads to an almost complete occupation swap from $n_2$ to $n_3$. Afterwards, the occupations of the dressed states remain approximately constant, however, the character of the dressed states changes, such that $|2\rangle$ is associated with the dark exciton, while $|3\rangle$ is a mixture between the ground state and bright exciton. In addition, a coherence between the dressed states is accumulated, resulting in the oscillation seen in the bare states. At the end of this phase a second anti-crossing appears, which results in another diabatic transition. 

This anti-crossing initializes the third phase. The magnitude of the population which returns from $|3\rangle$ to $|2\rangle$ depends on the one hand on the splitting between the eigenenergies and on the other hand on the relative phase, which the dressed states have accumulated since the first anti-crossing. For $\Theta=20\pi$ the phase relation favors the population of $|3\rangle$, resulting in a dark exciton occupation. In contrast, for $\Theta=21\pi$ the accumulated phase between the dressed states results in a destructive interference at the second anti-crossing, and the occupation $n_3$ drops down in favor of $n_2$. Hence the system returns to the ground state.

\subsection{Excitation with detuned pulses}

\begin{figure}[ht]{}
	\centering
	\includegraphics[width=1\columnwidth]{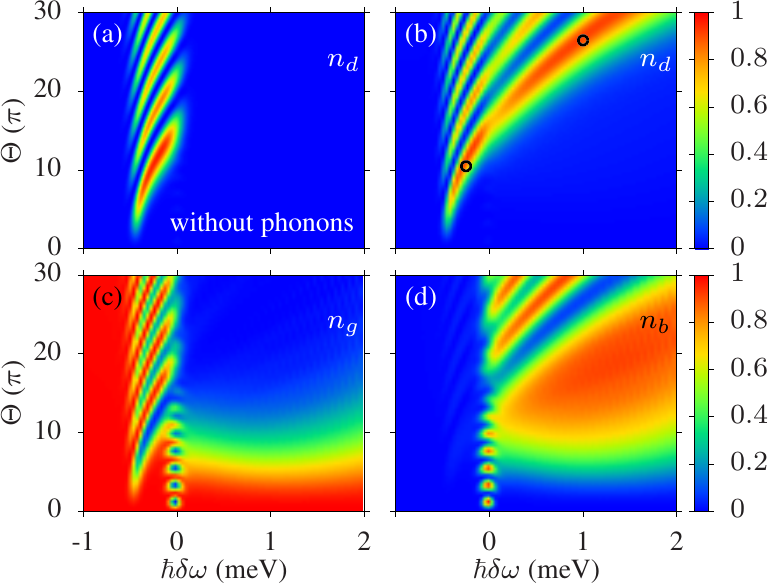}
      \caption{Final occupation of the QD states after excitation with a laser pulse with pulse area $\Theta$ and detuning $\delta\omega$. (a) Dark exciton occupation $n_d$ for the system without phonons. Occupation of (b) the dark exciton $n_d$, (c) the ground state $n_g$ and (d) the bright exciton $n_b$ including phonons. The small black circles refer to the parameters used in Fig.~\ref{fig6}.}
	\label{fig5}
\end{figure}
We now present a second method for the dark exciton preparation, which relies on the excitation with detuned laser pulses with fixed frequency. In this section the in-plane and out-of-plane component of the magnetic field are $B_x = 3\,{\rm{T}}$ and $B_z = 4\,{\rm{T}}$, resulting in a total field strength of $|\mathbf{B}| = 5\,{\rm{T}}$ with a tilt angle of $37^\circ$.

Fig.~\ref{fig5}(a) shows the final dark exciton occupation as a function of the pulse area $\Theta$ and the detuning $\delta\omega$ of the exciting laser pulse in the absence of phonons, the pulse duration is $\tau_0 = 12\,{\rm{ps}}$. Already without phonons a preparation of the dark exciton is possible for small negative detunings and we find a stripe pattern for detunings $-0.6\,{\rm{meV}} \lesssim \hbar\delta\omega \lesssim 0\,{\rm{meV}}$, while outside this region no dark exciton population is achieved. When the coupling to phonons is included, as shown in Fig.~\ref{fig5}(b), the dark exciton becomes accessible in a much larger parameter range. While the dark exciton preparation for negative detunings is almost unaffected by the phonons, the stripes are now extended into a region of positive detunings. The parameter range of positive detunings, which is promoted by the phonons, is much larger than the small region of negative detunings. Also here, we see oscillations, which for the negative detuning are with the ground state. Accordingly, the final occupation of the ground state in Fig.~\ref{fig5}(c) exhibits for negative detunings a stripe pattern, which is complementary to the dark exciton occupation. For positive chirps the oscillations occur between dark and bright states, such that the bright exciton occupation in (Fig.~\ref{fig5}(d)) shows the complementary stripe pattern for positive detunings. Finally we note that for large negative detuning with $\hbar\delta\omega < -0.6\,{\rm{meV}}$ the system remains completely in the ground state.

\begin{figure}[ht]{}
	\centering
	\includegraphics[width=1.0\columnwidth]{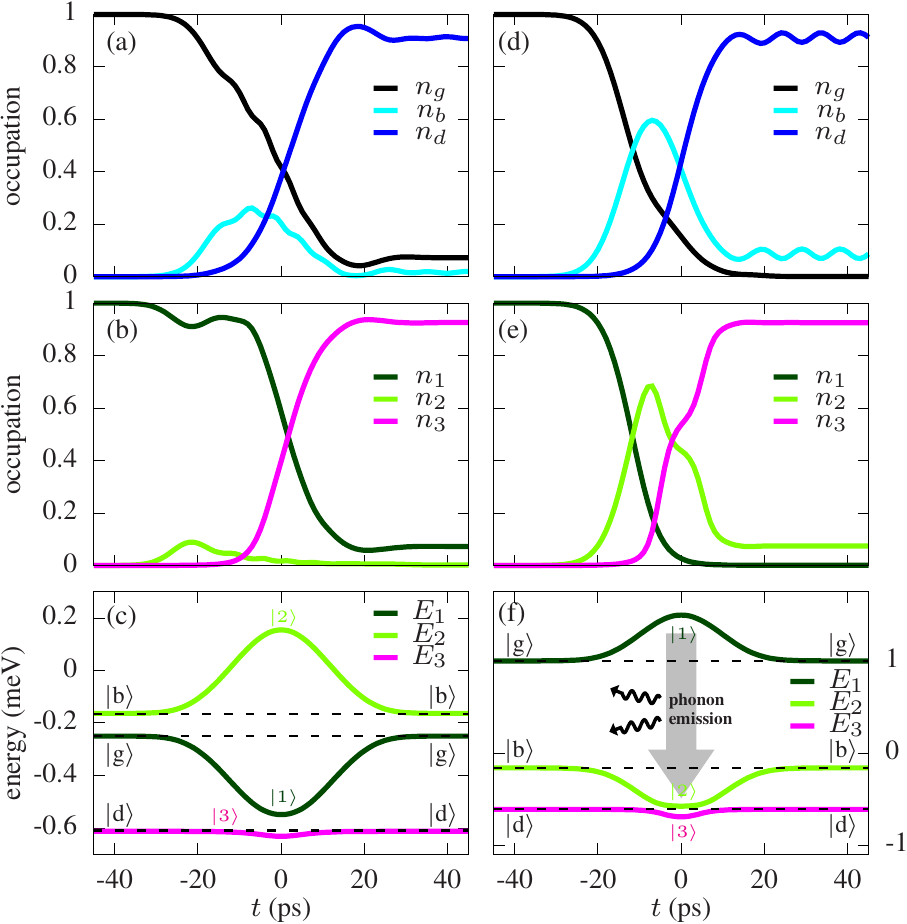}
      \caption{(a) Temporal evolution of the occupation of ground state $n_g$, bright exciton $n_b$, and dark exciton $n_d$ during the excitation with a laser pulse having a pulse area $\Theta = 10.5\pi$ and detuning $\hbar\delta\omega = -0.25\,{\rm{meV}}$. (b) Temporal evolution of the occupation of the dressed $n_i$ $(i=1,2,3)$. (c) Instantaneous eigenenergies; the dashed lines indicate the eigenenergies in absence of the coupling. (d), (e), and (f) show the same, but for $\hbar\delta\omega = 1\,{\rm{meV}}$ and $\Theta = 26.5\pi$.}
	\label{fig6}
\end{figure}

For this method the detuning is the crucial parameter. Thus, in Fig.~\ref{fig6} we study the excitation dynamics for two different detunings using again the dressed state picture. Figure~\ref{fig6}(a)-(c) shows the dynamics with a small negative detuning  $\hbar\delta\omega=-0.25\,{\rm{meV}}$ and Fig.~\ref{fig6}(d)-(f) is for a positive detuning with $\hbar\delta\omega=1.0\,{\rm{meV}}$. Figure~\ref{fig6}(a),(d) displays the temporal evolution of bare state occupations, Fig.~\ref{fig6}(b),(e) the dynamics of the dressed state occupations and in Fig.~\ref{fig6}(c),(f) the instantaneous eigenenergies are presented.

We first focus on the small negative detuning and consider a pulse area of $\Theta=10.5\pi$. These parameters correspond to the first maximum of the dark exciton occupation in the region of negative detunings in Fig.~\ref{fig5}(b), indicated by a small black circle. In the bare states dynamics we see that the initially occupied ground state is depleted successively during the pulse. While there is only a small transient occupation of the bright exciton, the population is almost completely transferred to the dark exciton. The population transfer occurs basically between two dressed states $|1\rangle$ and $|3\rangle$, while the state $|2\rangle$ remains nearly unaffected. The corresponding eigenenergies reveal, that this behavior is induced again by a diabatic transition between $|1\rangle$ and $|3\rangle$, inducing quantum beats. Please note the energy scale in Fig.~\ref{fig6}(c), which makes the eigenenergies $E_2$ and $E_3$ appear to be far away, but in fact they are quite close. Due to the negative detuning the eigenstate $|2\rangle$ has the highest energy, hence it does not take part in the dynamics, because also phonon-assisted transitions to the uppermost branch are suppressed at the low temperatures. Accordingly this excitation mechanism has the advantage that it omits the bright exciton, such that a deterioration due to the spontaneous decay is very unlikely.

The eigenenergies in Fig.~\ref{fig6}(c) also explain, why only for a small range of negative detunings a dark exciton preparation is possible. In absence of the laser pulse, the eigenenergy $E_2$ is fixed at the polaron energy $\hbar\omega_{pol} \approx -0.17\,{\rm{meV}}$. Similarly, the energy $E_3$, which is associated with the dark exciton, is only determined by the magnetic field and therefore fixed at $-0.6\,{\rm{meV}}$. In contrast, the position of the energy $E_1$ depends on the detuning, such that it lies at $-0.25\,{\rm{meV}}$ for the considered parameters. For larger detunings, this branch shifts to lower energies, until it lies below $E_3$ for $\hbar\delta\omega \lesssim -0.6\,{\rm{meV}}$. In this case, no anti-crossings with the other branches will occur, such that the system always remains on the lowest branch. Hence, for large negative detunings the population stays in the ground state.

Now we focus on the phonon-assisted dark exciton preparation using positively detuned laser pulses. Fig.~\ref{fig6}(d) shows the dynamics during the excitation with a $26.5\pi$-pulse, which is detuned by $1\,{\rm{meV}}$ from the polaron-shifted bright exciton transition. These parameters correspond to the first maximum of the dark exciton occupation in Fig.~\ref{fig5}(b), indicated by a black circle. The occupation of the initially occupied ground state $|g\rangle$ drops to zero when the laser pulse excites the system. After a transient occupation of the bright exciton $|b\rangle$, the population swaps eventually to the dark exciton $|d\rangle$. The dynamics of the dressed states (Fig.~\ref{fig6}(e)) differs from the case of negative detunings, since now all three eigenstates are involved in the excitation, which reveals a two-step process in analogy to the phonon-assisted dark exciton preparation using chirped pulses (cf. Fig.~\ref{fig4}). In the first step, the population of the initially occupied state $|1\rangle$ swaps to $|2\rangle$. In the second step, the transition to the eigenstate $|3\rangle$ occurs, which is associated with the dark exciton after the pulse. The eigenenergies (Fig.~\ref{fig6}(f)) show, that the order of the two uppermost branches is changed compared to the case of negative detuning, since now the highest energy corresponds to the ground state, while the intermediate branch is associated with the bright exciton. This arrangement promotes phonon-assisted transitions from the initially occupied state $|1\rangle$ to $|2\rangle$, while the state $|3\rangle$ is again not directly accessible due to the lack of a direct coupling between ground state and dark exciton. Around $t = 0\,{\rm{ps}}$ a wide anti-crossing region occurs, which promotes diabatic transitions from $|2\rangle$ to $|3\rangle$, that results in an occupation of the dark exciton. This also explains why, when varying the pulse area, the final state oscillates between bright and dark exciton.

\section{Conclusion}

In conclusion, we have presented two preparation protocols for the dark exciton in a single QD using a combination of a tilted magnetic field and a single laser pulse under the influence of phonons. The first method relies on the excitation with a chirped laser pulse. We have shown that the preparation mechanism using positive chirp coefficients is surprisingly stable against the impact of phonons. In addition, phonon-assisted processes promote a dark exciton preparation for negative chirps, which restores a symmetric situation of the dark exciton population with respect to the sign of the chirp.

The second method uses a detuned laser pulse. In analogy to the chirped laser pulses, the role of the phonons depends on the sign of the detuning. For negative detunings the dark exciton preparation does not rely on the mediation by phonons, but is also not impeded by the carrier-phonon coupling. For positive detuning the phonons have an active role in the preparation protocol, because the occupation of the dark exciton is accompanied with the emission of phonons. 

We have found that for the direct optical preparation schemes of the dark exciton proposed here, phonons are either not impeding the preparation or even helpful. In addition to the long lifetimes of the dark exciton, the extended parameter range  for the state preparation provided by the coupling to phonons makes the use of dark excitons even more attractive. Hence, we are confident, that our study will promote new experiments, exploring these fascinating states.


\begin{thebibliography}{32}

\makeatletter
\providecommand \@ifxundefined [1]{%
 \@ifx{#1\undefined}
}%
\providecommand \@ifnum [1]{%
 \ifnum #1\expandafter \@firstoftwo
 \else \expandafter \@secondoftwo
 \fi
}%
\providecommand \@ifx [1]{%
 \ifx #1\expandafter \@firstoftwo
 \else \expandafter \@secondoftwo
 \fi
}%
\providecommand \natexlab [1]{#1}%
\providecommand \enquote  [1]{``#1''}%
\providecommand \bibnamefont  [1]{#1}%
\providecommand \bibfnamefont [1]{#1}%
\providecommand \citenamefont [1]{#1}%
\providecommand \href@noop [0]{\@secondoftwo}%
\providecommand \href [0]{\begingroup \@sanitize@url \@href}%
\providecommand \@href[1]{\@@startlink{#1}\@@href}%
\providecommand \@@href[1]{\endgroup#1\@@endlink}%
\providecommand \@sanitize@url [0]{\catcode `\\12\catcode `\$12\catcode
  `\&12\catcode `\#12\catcode `\^12\catcode `\_12\catcode `\%12\relax}%
\providecommand \@@startlink[1]{}%
\providecommand \@@endlink[0]{}%
\providecommand \url  [0]{\begingroup\@sanitize@url \@url }%
\providecommand \@url [1]{\endgroup\@href {#1}{\urlprefix }}%
\providecommand \urlprefix  [0]{URL }%
\providecommand \Eprint [0]{\href }%
\providecommand \doibase [0]{http://dx.doi.org/}%
\providecommand \selectlanguage [0]{\@gobble}%
\providecommand \bibinfo  [0]{\@secondoftwo}%
\providecommand \bibfield  [0]{\@secondoftwo}%
\providecommand \translation [1]{[#1]}%
\providecommand \BibitemOpen [0]{}%
\providecommand \bibitemStop [0]{}%
\providecommand \bibitemNoStop [0]{.\EOS\space}%
\providecommand \EOS [0]{\spacefactor3000\relax}%
\providecommand \BibitemShut  [1]{\csname bibitem#1\endcsname}%
\let\auto@bib@innerbib\@empty
%</preamble>

\bibitem [{\citenamefont {Ramsay}\ \emph {et~al.}(2010)\citenamefont {Ramsay},
  \citenamefont {Gopal}, \citenamefont {Gauger}, \citenamefont {Nazir},
  \citenamefont {Lovett}, \citenamefont {Fox},\ and\ \citenamefont
  {Skolnick}}]{ramsay2010dam}%
  \BibitemOpen
  \bibfield  {author} {
  \bibinfo {author} {\bibfnamefont {A.~J.}\ \bibnamefont {Ramsay}},
  \bibinfo {author} {\bibfnamefont {A.~V.}\ \bibnamefont {Gopal}},
  \bibinfo {author} {\bibfnamefont {E.~M.}\ \bibnamefont {Gauger}},
  \bibinfo {author} {\bibfnamefont {A.}~\bibnamefont {Nazir}}, 
  \bibinfo {author} {\bibfnamefont {B.~W.}\ \bibnamefont {Lovett}},
  \bibinfo {author} {\bibfnamefont {A.~M.}\ \bibnamefont {Fox}}, \ and\
  \bibinfo {author} {\bibfnamefont {M.~S.}\ \bibnamefont {Skolnick}},\ }\bibfield  {title}
  {\enquote {\bibinfo {title} {{Damping of exciton rabi rotations by acoustic phonons in optically excited InGaAs/GaAs quantum dots}},}\ }\href@noop {}
  {\bibfield  {journal} {\bibinfo  {journal} {Phys. Rev. Lett.}\ }\textbf
  {\bibinfo {volume} {104}},\ \bibinfo {pages} {017402} (\bibinfo {year}
  {2010})}\BibitemShut {NoStop}%  
\bibitem [{\citenamefont {Reiter}\ \emph {et~al.}(2014)\citenamefont {Reiter},
  \citenamefont {Kuhn}, \citenamefont {Gl{\"a}ssl},\ and\ \citenamefont
  {Axt}}]{reiter2014the}%
  \BibitemOpen
  \bibfield  {author}{ 
  \bibinfo {author} {\bibfnamefont {D.~E.}\ \bibnamefont {Reiter}},
  \bibinfo {author} {\bibfnamefont {T.}~\bibnamefont {Kuhn}},
  \bibinfo {author} {\bibfnamefont {M.}~\bibnamefont {Gl{\"a}ssl}}, \ and\
  \bibinfo {author} {\bibfnamefont {V.~M.}\ \bibnamefont {Axt}},\ }\bibfield
  {title} {\enquote {\bibinfo {title} {The role of phonons for exciton and
  biexciton generation in an optically driven quantum dot},}\ }\href@noop {}
  {\bibfield  {journal} {\bibinfo {journal} {J. Phys. Condens. Matter}\ }\textbf
  {\bibinfo {volume} {26}},\ \bibinfo {pages} {423203} (\bibinfo {year}
  {2014})}\BibitemShut {NoStop}% 
\bibitem [{\citenamefont {Reiter}\ \emph {et~al.}(2012) \citenamefont{Reiter}, \citenamefont {L{\"u}ker}, \citenamefont {Gawarecki}, \citenamefont {Grodecka-Grad}, \citenamefont {Machnikowski}, \citenamefont {Axt},\ and\ \citenamefont {Kuhn}}]{reiter2012pho}%
  \BibitemOpen
  \bibfield  {author} {
  \bibinfo {author} {\bibfnamefont {D.~E.}\ \bibnamefont {Reiter}},    
  \bibinfo {author} {\bibfnamefont {S.}~\bibnamefont {L{\"u}ker}},
  \bibinfo {author} {\bibfnamefont {K.}~\bibnamefont {Gawarecki}}, 
  \bibinfo {author} {\bibfnamefont {A.}~\bibnamefont {Grodecka-Grad}},
  \bibinfo {author} {\bibfnamefont {P.}~\bibnamefont {Machnikowski}},
  \bibinfo {author} {\bibfnamefont {V.~M.}\ \bibnamefont {Axt}}, \ and\
  \bibinfo {author} {\bibfnamefont {T.}~\bibnamefont {Kuhn}},\ }
  \bibfield  {title} {\enquote
  {\bibinfo {title} {Phonon effects on population inversion in quantum dots: Resonant, detuned and frequency-swept excitations},}\ }
  \href@noop {} {\ \bibinfo {journal}{Acta Phys. Pol. A} \textbf {\bibinfo   {volume} {122}},\ \bibinfo {pages} {1065} (\bibinfo {year}
  {2012})}\BibitemShut {NoStop}% 
\bibitem [{\citenamefont {Gl\"assl}\ \emph {et~al.}(2013)\citenamefont {Gl\"assl}, \citenamefont {Barth},\ and\ \citenamefont {Axt}}]{glassl2013pro}%
  \BibitemOpen
  \bibfield  {author} {
  \bibinfo {author} {\bibfnamefont {M.}~\bibnamefont {Gl\"assl}}, 
  \bibinfo {author} {\bibfnamefont {A.~M.}\ \bibnamefont {Barth}},\ and\
  \bibinfo {author} {\bibfnamefont {V.~M.}\ \bibnamefont {Axt}},\
  }\bibfield  {title} {\enquote {\bibinfo {title} {Proposed robust and
  high-fidelity preparation of excitons and biexcitons in semiconductor quantum dots making active use of phonons},}\ }
  \href@noop {} {\bibfield  {journal}
  {\bibinfo  {journal} {Phys. Rev. Lett.}\ }\textbf {\bibinfo {volume} {110}},\
  \bibinfo {pages} {147401} (\bibinfo {year} {2013})}\BibitemShut {NoStop}%
\bibitem [{\citenamefont {Bounouar}\ \emph {et~al.}(2015)\citenamefont {Bounouar}, \citenamefont {M\"uller}, \citenamefont {Barth}, \citenamefont
  {Gl\"assl}, \citenamefont {Axt},\ and\ \citenamefont {Michler}}]{bounouar2015pho}%
  \BibitemOpen
  \bibfield  {author} {
  \bibinfo {author} {\bibfnamefont {S.}~\bibnamefont {Bounouar}}, 
  \bibinfo {author} {\bibfnamefont {M.}~\bibnamefont {M\"uller}},
  \bibinfo {author} {\bibfnamefont {A.~M.}\ \bibnamefont {Barth}}, 
  \bibinfo {author} {\bibfnamefont {M.}~\bibnamefont {Gl\"assl}}, 
  \bibinfo {author} {\bibfnamefont {V.~M.}\ \bibnamefont {Axt}}, \ and\
  \bibinfo {author} {\bibfnamefont {P.}~\bibnamefont {Michler}},\ }\bibfield  {title} {\enquote
  {\bibinfo {title} {Phonon-assisted robust and deterministic two-photon
  biexciton preparation in a quantum dot},}\ }\href@noop {} 
  {\bibfield  {journal} {\bibinfo  {journal} {Phys. Rev. B}\ }\textbf {\bibinfo {volume}
  {91}},\ \bibinfo {pages} {161302} (\bibinfo {year} {2015})}\BibitemShut
  {NoStop}%
\bibitem [{\citenamefont {Ardelt}\ \emph {et~al.}(2014)\citenamefont {Ardelt}, \citenamefont {Hanschke}, \citenamefont {Fischer}, \citenamefont {M\"uller}, \citenamefont {Kleinkauf}, \citenamefont {Koller}, \citenamefont {Bechtold}, \citenamefont {Simmet}, \citenamefont {Wierzbowski}, \citenamefont {Riedl}, \citenamefont {Abstreiter},\ and\ \citenamefont {Finley}}]{ardelt2014dis}%
  \BibitemOpen \bibfield  {author} {
  \bibinfo {author} {\bibfnamefont {P.-L.}\ \bibnamefont {Ardelt}},
  \bibinfo {author} {\bibfnamefont {L.}~\bibnamefont {Hanschke}},
  \bibinfo {author} {\bibfnamefont {K.~A.}\ \bibnamefont {Fischer}},
  \bibinfo {author} {\bibfnamefont {K.}~\bibnamefont {M\"uller}}, 
  \bibinfo {author} {\bibfnamefont {A.}~\bibnamefont {Kleinkauf}}, 
  \bibinfo {author} {\bibfnamefont {M.}~\bibnamefont {Koller}}, 
  \bibinfo {author} {\bibfnamefont {A.}~\bibnamefont {Bechtold}}, 
  \bibinfo {author} {\bibfnamefont {T.}~\bibnamefont {Simmet}}, 
  \bibinfo {author} {\bibfnamefont {J.}~\bibnamefont {Wierzbowski}},
  \bibinfo {author} {\bibfnamefont {H.}~\bibnamefont {Riedl}}, 
  \bibinfo {author} {\bibfnamefont {G.}~\bibnamefont {Abstreiter}}, \ and\
  \bibinfo {author} {\bibfnamefont {J.~J.}\ \bibnamefont {Finley}},\ }\bibfield  {title} {\enquote {\bibinfo
  {title} {Dissipative preparation of the exciton and biexciton in
  self-assembled quantum dots on picosecond time scales},}\ }\href@noop {}
  {\bibfield  {journal} 
  {\bibinfo  {journal} {Phys. Rev. B}\ }\textbf {\bibinfo
  {volume} {90}},\ \bibinfo {pages} {241404} (\bibinfo {year}
  {2014})}\BibitemShut {NoStop}%  
\bibitem [{\citenamefont {Quilter}\ \emph {et~al.}(2015)\citenamefont
  {Quilter}, \citenamefont {Brash}, \citenamefont {Liu}, \citenamefont
  {Gl\"assl}, \citenamefont {Barth}, \citenamefont {Axt}, \citenamefont
  {Ramsay}, \citenamefont {Skolnick},\ and\ \citenamefont
  {Fox}}]{quilter2015pho}%
  \BibitemOpen
  \bibfield  {author} {
  \bibinfo {author} {\bibfnamefont {J.~H.}\ \bibnamefont {Quilter}},
  \bibinfo {author} {\bibfnamefont {A.~J.}\ \bibnamefont {Brash}},
  \bibinfo {author} {\bibfnamefont {F.}~\bibnamefont {Liu}}, 
  \bibinfo {author} {\bibfnamefont {M.}~\bibnamefont {Gl\"assl}}, 
  \bibinfo {author} {\bibfnamefont {A.~M.}\ \bibnamefont {Barth}}, 
  \bibinfo {author} {\bibfnamefont {V.~M.}\ \bibnamefont {Axt}}, 
  \bibinfo {author} {\bibfnamefont {A.~J.}\ \bibnamefont {Ramsay}},
  \bibinfo {author} {\bibfnamefont {M.~S.}\ \bibnamefont {Skolnick}}, \ and\ \bibinfo {author} {\bibfnamefont {A.~M.}\ \bibnamefont {Fox}},\ }\bibfield  {title} {\enquote {\bibinfo {title}
  {Phonon-assisted population inversion of a single
  $\mathrm{InGaAs}/\mathrm{GaAs}$ quantum dot by pulsed laser excitation},}\
  }\href@noop {} {\bibfield  {journal} 
  {\bibinfo  {journal} {Phys. Rev. Lett.}\ }\textbf {\bibinfo {volume} {114}},\ \bibinfo {pages} {137401} (\bibinfo
  {year} {2015})}\BibitemShut {NoStop}%   
\bibitem [{\citenamefont {Manson}\ \emph {et~al.}(2013)\citenamefont {Manson}, \citenamefont {Roy-Choudhury},\ and\ \citenamefont {Hughes}}]{manson2016pol}%
  \BibitemOpen
  \bibfield  {author} {
  \bibinfo {author} {\bibfnamefont {R.}~\bibnamefont {Manson}}, 
  \bibinfo {author} {\bibfnamefont {K.}~\bibnamefont {Roy-Choudhury}},\ and\
  \bibinfo {author} {\bibfnamefont {S.}~\bibnamefont {Hughes}},\
  }\bibfield  {title} {\enquote {\bibinfo {title} {Polaron master equation theory of pulse-driven phonon-assisted population inversion and single-photon emission from quantum-dot excitons},}\ }
  \href@noop {} {\bibfield  {journal}
  {\bibinfo  {journal} {Phys. Rev. B}\ }\textbf {\bibinfo {volume} {93}},\
  \bibinfo {pages} {155423} (\bibinfo {year} {2016})}\BibitemShut {NoStop}%    
\bibitem [{\citenamefont {Barth}\ \emph {et~al.}(2016)\citenamefont {Barth}, \citenamefont {L{\"u}ker}, \citenamefont {Vagov}, \citenamefont {Reiter},  \citenamefont {Kuhn},\ and\ \citenamefont {Axt}}]{barth2016fas}%
  \BibitemOpen
  \bibfield  {author} {
  \bibinfo {author} {\bibfnamefont {A.~M.}\ \bibnamefont {Barth}}, 
  \bibinfo {author} {\bibfnamefont {S.}~\bibnamefont {L{\"u}ker}},
  \bibinfo {author} {\bibfnamefont {A.}~\bibnamefont {Vagov}}, 
  \bibinfo {author} {\bibfnamefont {D.~E.}\ \bibnamefont {Reiter}}, 
  \bibinfo {author} {\bibfnamefont {T.}~\bibnamefont {Kuhn}}, \ and\
  \bibinfo {author} {\bibfnamefont {V.~M.}\ \bibnamefont {Axt}},\ }\bibfield  {title} {\enquote
  {\bibinfo {title} {Fast and selective phonon-assisted state preparation of a quantum dot by adiabatic undressing},}\ }\href@noop {} {\bibfield  {journal}  {\bibinfo {journal} {Phys. Rev. B}\ }\textbf {\bibinfo
  {volume} {94}},\ \bibinfo {pages} {045306} (\bibinfo {year}
  {2016})}\BibitemShut {NoStop}%  
\bibitem [{\citenamefont {Liu}\ \emph {et~al.}(2016)\citenamefont {Liu},
  \citenamefont {Martins}, \citenamefont {Brash}, \citenamefont {Barth},
  \citenamefont {Quilter}, \citenamefont {Axt}, \citenamefont {Skolnick},\ and\ \citenamefont {Fox}}]{liu2016ult}%
  \BibitemOpen
  \bibfield  {author} {
  \bibinfo {author} {\bibfnamefont {F.}~\bibnamefont {Liu}}, 
  \bibinfo {author} {\bibfnamefont {L.~M.~P.}\ \bibnamefont {Martins}},
  \bibinfo {author} {\bibfnamefont {A.~J.}\ \bibnamefont {Brash}}, 
  \bibinfo {author} {\bibfnamefont {A.~M.}\ \bibnamefont {Barth}}, 
  \bibinfo {author} {\bibfnamefont {J.~H.}\ \bibnamefont {Quilter}},
  \bibinfo {author} {\bibfnamefont {V.~M.}\ \bibnamefont {Axt}}, 
  \bibinfo {author} {\bibfnamefont {M.~S.}\ \bibnamefont {Skolnick}}, \ and\ \bibinfo {author} {\bibfnamefont {A.~M.}\ \bibnamefont {Fox}},\ }\bibfield  {title} {\enquote {\bibinfo {title}
  {Ultrafast depopulation of a quantum dot by $\mathrm{LA}$-phonon-assisted stimulated
  emission},}\ }\href@noop {} 
  {\bibfield  {journal} 
  {\bibinfo {journal} {Phys. Rev. B}\ }\textbf{\bibinfo {volume} {93}},\ \bibinfo {pages} {161407} (\bibinfo {year}
  {2016})}\BibitemShut {NoStop}%  
\bibitem [{\citenamefont {Stevenson}\ \emph {et~al.}(2004)\citenamefont
  {Stevenson}, \citenamefont {Young}, \citenamefont {See}, \citenamefont
  {Farrer}, \citenamefont {Ritchie},\ and\ \citenamefont
  {Shields}}]{stevenson2004tim}%
  \BibitemOpen
  \bibfield  {author} {
  \bibinfo {author} {\bibfnamefont {R.~M.}\ \bibnamefont {Stevenson}},
  \bibinfo {author} {\bibfnamefont {R.~J.}\ \bibnamefont {Young}}, 
  \bibinfo {author} {\bibfnamefont {P.}~\bibnamefont {See}}, 
  \bibinfo {author} {\bibfnamefont {I.}~\bibnamefont {Farrer}}, 
  \bibinfo {author} {\bibfnamefont {D.~A.}\ \bibnamefont {Ritchie}}, \ and\
   \bibinfo {author} {\bibfnamefont {A.~J.}\ \bibnamefont {Shields}},\ }\bibfield  {title}
  {\enquote {\bibinfo {title} {Time-resolved studies of single quantum dots in magnetic fields},}\ }\href@noop {} 
  {\bibfield  {journal} {\bibinfo  {journal}
  {Physica E}\ }\textbf {\bibinfo
  {volume} {21}},\ \bibinfo {pages} {381} (\bibinfo {year} {2004})}\BibitemShut
  {NoStop}%
\bibitem [{\citenamefont {Korkusinski}\ and\ \citenamefont
  {Hawrylak}(2013)}]{korkusinski2013ato}%
  \BibitemOpen
  \bibfield  {author} {
  \bibinfo {author} {\bibfnamefont {M.}~\bibnamefont {Korkusinski}}\ and\ \bibinfo {author} {\bibfnamefont {P.}~\bibnamefont {Hawrylak}},\ }
  \bibfield  {title} {\enquote {\bibinfo {title} {Atomistic
  theory of emission from dark excitons in self-assembled quantum dots},}\
  }\href@noop {} {\bibfield  {journal} {\bibinfo  {journal} {Phys. Rev. B}\
  }\textbf {\bibinfo {volume} {87}},\ \bibinfo {pages} {115310} (\bibinfo
  {year} {2013})}\BibitemShut {NoStop}%
\bibitem [{\citenamefont {Poem}\ \emph {et~al.}(2010)\citenamefont {Poem},
  \citenamefont {Kodriano}, \citenamefont {Tradonsky}, \citenamefont {Lindner},
  \citenamefont {Gerardot}, \citenamefont {Petroff},\ and\ \citenamefont
  {Gershoni}}]{poem2010acc}%
  \BibitemOpen
  \bibfield  {author} {
  \bibinfo {author} {\bibfnamefont {E.}~\bibnamefont {Poem}}, 
  \bibinfo {author} {\bibfnamefont {Y.}~\bibnamefont {Kodriano}},
  \bibinfo {author} {\bibfnamefont {C.}~\bibnamefont {Tradonsky}}, 
  \bibinfo {author} {\bibfnamefont {N.~H.}\ \bibnamefont {Lindner}},
  \bibinfo {author} {\bibfnamefont {B.~D.}\ \bibnamefont {Gerardot}},
  \bibinfo {author} {\bibfnamefont {P.~M.}\ \bibnamefont {Petroff}}, \ and\
  \bibinfo {author} {\bibfnamefont {D.}~\bibnamefont {Gershoni}},\ }\bibfield  {title} {\enquote
  {\bibinfo {title} {Accessing the dark exciton with light},}\ }\href@noop {}
 {\bibfield  {journal} {\bibinfo  {journal} {Nat. Physics}\
  }\textbf {\bibinfo {volume} {6}},\ \bibinfo {pages} {993} (\bibinfo
  {year} {2010})}\BibitemShut {NoStop}%
\bibitem [{\citenamefont {Smole\'{n}ski}\ \emph {et~al.}(2015)\citenamefont
  {Smole\'{n}ski}, \citenamefont {Kazimierczuk}, \citenamefont {Goryca},
  \citenamefont {Wojnar},\ and\ \citenamefont {Kossacki}}]{smolenski2015mec}%
  \BibitemOpen
  \bibfield  {author} {
  \bibinfo {author} {\bibfnamefont {T.}~\bibnamefont {Smole\'{n}ski}},
  \bibinfo {author} {\bibfnamefont {T.}~\bibnamefont {Kazimierczuk}},
  \bibinfo {author} {\bibfnamefont {M.}~\bibnamefont {Goryca}}, 
  \bibinfo {author} {\bibfnamefont {P.}~\bibnamefont {Wojnar}}, \
  and\ \bibinfo {author} {\bibfnamefont {P.}~\bibnamefont {Kossacki}},\
  }\bibfield  {title} {\enquote {\bibinfo {title} {Mechanism and dynamics of biexciton formation from a long-lived dark exciton in a $\mathrm{CdTe}$ quantum dot},}\
  }\href@noop {} {\bibfield  {journal} {\bibinfo  {journal} {Phys. Rev. B}\
  }\textbf {\bibinfo {volume} {91}},\ \bibinfo {pages} {155430} (\bibinfo
  {year} {2015})}\BibitemShut {NoStop}%  
\bibitem [{\citenamefont {Schmidgall}\ \emph {et~al.}(2015)\citenamefont
  {Schmidgall}, \citenamefont {Schwartz}, \citenamefont {Cogan}, \citenamefont
  {Gantz}, \citenamefont {Heindel}, \citenamefont {Reitzenstein},\ and\
  \citenamefont {Gershoni}}]{schmidtgall2015all}%
  \BibitemOpen
  \bibfield  {author} {
  \bibinfo {author} {\bibfnamefont {E.~R.}\ \bibnamefont {Schmidgall}},
  \bibinfo {author} {\bibfnamefont {D.}~\bibnamefont {Schwartz}}, 
  \bibinfo {author} {\bibfnamefont {D.}~\bibnamefont {Cogan}},
  \bibinfo {author} {\bibfnamefont {L.}~\bibnamefont {Gantz}}, 
  \bibinfo {author} {\bibfnamefont {T.}~\bibnamefont {Heindel}}, 
  \bibinfo {author} {\bibfnamefont {S.}~\bibnamefont {Reitzenstein}}, \ and\ \bibinfo {author} {\bibfnamefont {D.}~\bibnamefont {Gershoni}},\ }\bibfield  {title} {\enquote
  {\bibinfo {title} {All-optical depletion of dark excitons from a
  semiconductor quantum dot},}\ }\href@noop {} {\bibfield  {journal} {\bibinfo
  {journal} {Appl. Phys. Lett.}\ }\textbf {\bibinfo {volume} {106}},\
  \bibinfo {pages} {193101} (\bibinfo {year} {2015})}\BibitemShut {NoStop}%  
\bibitem [{\citenamefont {Schwartz}\ \emph {et~al.}(2015)\citenamefont
  {Schwartz}, \citenamefont {Schmidgall}, \citenamefont {Gantz}, \citenamefont
  {Cogan}, \citenamefont {Bordo}, \citenamefont {Don}, \citenamefont
  {Zielinski},\ and\ \citenamefont {Gershoni}}]{schwartz2015det}%
  \BibitemOpen
  \bibfield  {author} {
  \bibinfo {author} {\bibfnamefont {I.}~\bibnamefont {Schwartz}}, 
  \bibinfo {author} {\bibfnamefont {E.~R.}~\bibnamefont {Schmidgall}},
  \bibinfo {author} {\bibfnamefont {L.}~\bibnamefont {Gantz}},
  \bibinfo {author} {\bibfnamefont {D.}~\bibnamefont {Cogan}}, 
  \bibinfo {author} {\bibfnamefont {E.}~\bibnamefont {Bordo}}, 
  \bibinfo {author} {\bibfnamefont {Y.}~\bibnamefont {Don}}, 
  \bibinfo {author} {\bibfnamefont {M.}~\bibnamefont {Zielinski}}, \ and\
  \bibinfo {author} {\bibfnamefont {D.}~\bibnamefont {Gershoni}},\ }\bibfield  {title} {\enquote {\bibinfo {title} {Deterministic
  writing and control of the dark exciton spin using single short optical
  pulses},}\ }\href@noop {} 
  {\bibfield  {journal} {\bibinfo  {journal} {Phys. Rev. X}\
  }\textbf {\bibinfo {volume} {5}},\ \bibinfo {pages} {011009} (\bibinfo
  {year} {2015})}\BibitemShut {NoStop}%  
\bibitem [{\citenamefont {L{\"u}ker}\ \emph {et~al.}(2015)\citenamefont
  {L{\"u}ker}, \citenamefont {Kuhn},\ and\ \citenamefont
  {Reiter}}]{luker2015dir}%
  \BibitemOpen
  \bibfield  {author} {
  \bibinfo {author} {\bibfnamefont {S.}~\bibnamefont {L{\"u}ker}}, 
  \bibinfo {author} {\bibfnamefont {T.}~\bibnamefont {Kuhn}}, \ and\
  \bibinfo {author} {\bibfnamefont {D.~E.}\ \bibnamefont {Reiter}},\
  }\bibfield  {title} {\enquote {\bibinfo {title} {Direct optical state
  preparation of the dark exciton in a quantum dot},}\ }\href@noop {}
  {\bibfield  {journal} {\bibinfo  {journal} {Phys. Rev. B}\
  }\textbf {\bibinfo {volume} {92}},\ \bibinfo {pages} {201305} (\bibinfo
  {year} {2015})}\BibitemShut {NoStop}%  
\bibitem [{\citenamefont {Vagov}\ \emph {et~al.}(2004)\citenamefont {Vagov},
  \citenamefont {Axt}, \citenamefont {Kuhn}, \citenamefont {Langbein},
  \citenamefont {Borri},\ and\ \citenamefont {Woggon}}]{woggon2004non}%
  \BibitemOpen
  \bibfield  {author} {
  \bibinfo {author} {\bibfnamefont {A.}~\bibnamefont {Vagov}}, 
  \bibinfo {author} {\bibfnamefont {V.~M.}\ \bibnamefont {Axt}},
  \bibinfo {author} {\bibfnamefont {T.}~\bibnamefont {Kuhn}}, 
  \bibinfo {author} {\bibfnamefont {W.}~\bibnamefont {Langbein}}, 
  \bibinfo {author} {\bibfnamefont {P.}~\bibnamefont {Borri}}, \ and\
  \bibinfo {author} {\bibfnamefont {U.}~\bibnamefont {Woggon}},\ }\bibfield  {title} {\enquote
  {\bibinfo {title} {Nonmonotonous temperature dependence of the initial
  decoherence in quantum dots},}\ }\href@noop {} {\bibfield  {journal}
  {\bibinfo  {journal} {Phys. Rev. B}\ }\textbf {\bibinfo {volume} {70}},\
  \bibinfo {pages} {201305} (\bibinfo {year} {2004})}\BibitemShut {NoStop}% 
\bibitem [{\citenamefont {Krummheuer}\ \emph {et~al.}(2002)\citenamefont
  {Krummheuer}, \citenamefont {Axt},\ and\ \citenamefont
  {Kuhn}}]{krummheuer2002the}%
  \BibitemOpen
  \bibfield  {author} {
  \bibinfo {author} {\bibfnamefont {B.}~\bibnamefont {Krummheuer}},
  \bibinfo {author} {\bibfnamefont {V.~M.}\ \bibnamefont {Axt}},
  \ and\ \bibinfo {author} {\bibfnamefont {T.}~\bibnamefont {Kuhn}},\
  }\bibfield  {title} {\enquote {\bibinfo {title} {Theory of pure dephasing and the resulting absorption line shape in semiconductor quantum dots},}\
  }\href@noop {} {\bibfield  {journal} {\bibinfo  {journal} {Phys. Rev. B}\
  }\textbf {\bibinfo {volume} {65}},\ \bibinfo {pages} {195313} (\bibinfo
  {year} {2002})}\BibitemShut {NoStop}%
\bibitem [{\citenamefont {Kr{\"u}gel}\ \emph {et~al.}(2006)\citenamefont
  {Kr{\"u}gel}, \citenamefont {Axt},\ and\ \citenamefont
  {Kuhn}}]{krugel2006bac}%
  \BibitemOpen
  \bibfield  {author} {
  \bibinfo {author} {\bibfnamefont {A.}~\bibnamefont {Kr{\"u}gel}},
  \bibinfo {author} {\bibfnamefont {V.~M.}\ \bibnamefont {Axt}},
  \ and\ \bibinfo {author} {\bibfnamefont {T.}~\bibnamefont {Kuhn}},\
  }\bibfield  {title} {\enquote {\bibinfo {title} {Back action of
  nonequilibrium phonons on the optically induced dynamics in semiconductor
  quantum dots},}\ }\href@noop {} {\bibfield  {journal} {\bibinfo  {journal}
  {Phys. Rev. B}\ }\textbf {\bibinfo {volume} {73}},\ \bibinfo {pages} {035302}
  (\bibinfo {year} {2006})}\BibitemShut {NoStop}%
\bibitem [{\citenamefont {L{\"u}ker}\ \emph {et~al.}(2012)\citenamefont
  {L{\"u}ker}, \citenamefont {Gawarecki}, \citenamefont {Reiter}, \citenamefont
  {Grodecka-Grad}, \citenamefont {Axt}, \citenamefont {Machnikowski},\ and\
  \citenamefont {Kuhn}}]{luker2012inf}%
  \BibitemOpen
\bibfield  {journal} {  }\bibfield  {author} {
  \bibinfo {author} {\bibfnamefont {S.}~\bibnamefont {L{\"u}ker}}, 
  \bibinfo {author} {\bibfnamefont {K.}~\bibnamefont {Gawarecki}}, 
  \bibinfo {author} {\bibfnamefont {D.~E.}\ \bibnamefont {Reiter}},
  \bibinfo {author} {\bibfnamefont {A.}~\bibnamefont {Grodecka-Grad}},
  \bibinfo {author} {\bibfnamefont {V.~M.}\ \bibnamefont {Axt}}, 
  \bibinfo {author} {\bibfnamefont {P.}~\bibnamefont {Machnikowski}}, \
  and\ \bibinfo {author} {\bibfnamefont {T.}~\bibnamefont {Kuhn}},\ }\bibfield
  {title} {\enquote {\bibinfo {title} {Influence of acoustic phonons on the
  optical control of quantum dots driven by adiabatic rapid passage},}\
  }\href@noop {} {\bibfield  {journal} {\bibinfo  {journal} {Phys. Rev. B}\
  }\textbf {\bibinfo {volume} {85}},\ \bibinfo {pages} {121302} (\bibinfo
  {year} {2012})}\BibitemShut {NoStop}%
\bibitem [{\citenamefont {Kaldewey}\ \emph {et~al.}()\citenamefont {Kaldewey},
  \citenamefont {L{\"u}ker}, \citenamefont {Kuhlmann}, \citenamefont {Valentin},
  \citenamefont {Ludwig}, \citenamefont {Wieck}, \citenamefont {Reiter},
  \citenamefont {Kuhn},\ and\ \citenamefont {Warburton}}]{kaldewey2016dem}%
  \BibitemOpen
  \bibfield  {author} {
  \bibinfo {author} {\bibfnamefont {T.}~\bibnamefont {Kaldewey}}, 
  \bibinfo {author} {\bibfnamefont {S.}~\bibnamefont {L{\"u}ker}},
  \bibinfo {author} {\bibfnamefont {A.~V.}\ \bibnamefont {Kuhlmann}},
  \bibinfo {author} {\bibfnamefont {S.~R.}\ \bibnamefont {Valentin}},
  \bibinfo {author} {\bibfnamefont {A.}~\bibnamefont {Ludwig}}, 
  \bibinfo {author} {\bibfnamefont {A.~D.}\ \bibnamefont {Wieck}}, 
  \bibinfo {author} {\bibfnamefont {D.~E.}\ \bibnamefont {Reiter}},
  \bibinfo {author} {\bibfnamefont {T.}~\bibnamefont {Kuhn}}, \ and\
  \bibinfo {author} {\bibfnamefont {R.}~\bibnamefont {Warburton}},\ }\bibfield  {title} {\enquote {\bibinfo {title} {Demonstrating
  the decoupling regime of the electron-phonon interaction in a quantum dot
  using chirped optical excitation},}\ }\href@noop {} {\bibinfo  {journal}
  arXiv preprint arXiv:1701.01304\ }\BibitemShut {NoStop}%       
\bibitem [{\citenamefont {Kaldewey}\ \emph {et~al.}()\citenamefont {Kaldewey},
  \citenamefont {L{\"u}ker}, \citenamefont {Kuhlmann}, \citenamefont {Valentin},
  \citenamefont {Ludwig}, \citenamefont {Wieck}, \citenamefont {Reiter},
  \citenamefont {Kuhn},\ and\ \citenamefont {Warburton}}]{kaldewey2017coh}%
  \BibitemOpen
  \bibfield  {author} {
  \bibinfo {author} {\bibfnamefont {T.}~\bibnamefont {Kaldewey}}, 
  \bibinfo {author} {\bibfnamefont {S.}~\bibnamefont {L{\"u}ker}},
  \bibinfo {author} {\bibfnamefont {A.~V.}\ \bibnamefont {Kuhlmann}},
  \bibinfo {author} {\bibfnamefont {S.~R.}\ \bibnamefont {Valentin}},
  \bibinfo {author} {\bibfnamefont {A.}~\bibnamefont {Ludwig}}, 
  \bibinfo {author} {\bibfnamefont {A.~D.}\ \bibnamefont {Wieck}}, 
  \bibinfo {author} {\bibfnamefont {D.~E.}\ \bibnamefont {Reiter}},
  \bibinfo {author} {\bibfnamefont {T.}~\bibnamefont {Kuhn}}, \ and\
  \bibinfo {author} {\bibfnamefont {R.}~\bibnamefont {Warburton}},\ }\bibfield  {title} {\enquote {\bibinfo {title} {Coherent and robust high-fidelity generation of a biexciton in a quantum dot by rapid adiabatic passage},}\ }\href@noop {} {\bibfield  {journal} {\bibinfo  {journal} {Phys. Rev. B}\
  }\textbf {\bibinfo {volume} {95}},\ \bibinfo {pages} {161302} (\bibinfo
  {year} {2017})}\BibitemShut {NoStop}%       
\bibitem [{\citenamefont {Gl{\"a}ssl}\ \emph {et~al.}(2011)\citenamefont
  {Gl{\"a}ssl}, \citenamefont {Vagov}, \citenamefont {L{\"u}ker}, \citenamefont {Reiter}, \citenamefont {Croitoru}, \citenamefont {Machnikowski}, \citenamefont {Axt}, \ and\ \citenamefont
  {Kuhn}}]{glassl2011lon}%
  \BibitemOpen
  \bibfield  {author} {
  \bibinfo {author} {\bibfnamefont {M.}~\bibnamefont {Gl{\"a}ssl}},
  \bibinfo {author} {\bibfnamefont {A.}~ \bibnamefont {Vagov}},
  \bibinfo {author} {\bibfnamefont {S.}~ \bibnamefont {L{\"u}ker}},  
  \bibinfo {author} {\bibfnamefont {D.~E.}\ \bibnamefont {Reiter}},
  \bibinfo {author} {\bibfnamefont {M.~D.}\ \bibnamefont {Croitoru}},
  \bibinfo {author} {\bibfnamefont {P.}~ \bibnamefont {Machnikowski}},
  \bibinfo {author} {\bibfnamefont {V.~M.}\ \bibnamefont {Axt}},      
  \ and\ \bibinfo {author} {\bibfnamefont {T.}~\bibnamefont {Kuhn}},\
  }\bibfield  {title} {\enquote {\bibinfo {title} {Long-time dynamics and stationary nonequilibrium of an optically driven strongly confined quantum dot coupled to phonons},}\ }\href@noop {} {\bibfield  {journal} {\bibinfo  {journal}
  {Phys. Rev. B}\ }\textbf {\bibinfo {volume} {84}},\ \bibinfo {pages} {195311}
  (\bibinfo {year} {2011})}\BibitemShut {NoStop}%     
\bibitem [{\citenamefont {Landau}(1932)}]{landau1932ath}%
  \BibitemOpen
  \bibfield  {author} {\bibinfo {author} {\bibfnamefont {L.~D.}\ \bibnamefont
  {Landau}},\ }\bibfield  {title} {\enquote {\bibinfo {title} {A theory of
  energy transfer on collisions.}}\ }\href@noop {} {\bibfield  {journal}
  {\bibinfo  {journal} {Phys. Z. Sowjet.}\ }\textbf {\bibinfo {volume} {1}},\
  \bibinfo {pages} {88} (\bibinfo {year} {1932})}\BibitemShut {NoStop}%
\bibitem [{\citenamefont {Zener}(1932)}]{zener1932non}%
  \BibitemOpen
  \bibfield  {author} {
  \bibinfo {author} {\bibfnamefont {C.}\ \bibnamefont {Zener}},\ }\bibfield  {title} {\enquote {\bibinfo {title}
  {Non-adiabatic crossing of energy levels},}\ }\href@noop {} {\bibfield
  {journal} {\bibinfo  {journal} {Proc. R. Soc. A}\ }\textbf {\bibinfo {volume}
  {137}},\ \bibinfo {pages} {696} (\bibinfo {year} {1932})}\BibitemShut
  {NoStop}%  
\bibitem [{\citenamefont {Tannor}(2007)}]{tannor2007int}%
  \BibitemOpen
  \bibfield  {author} {\bibinfo {author} {\bibfnamefont {D.~J.}\ \bibnamefont
  {Tannor}},\ }\href@noop {} {\emph {\bibinfo {title} {Introduction to quantum
  mechanics}}}\ (\bibinfo  {publisher} {University Science Books},\ \bibinfo
  {address} {Sausalito, California},\ \bibinfo {year} {2007})\BibitemShut
  {NoStop}%
\bibitem [{\citenamefont {Wu}\ \emph {et~al.}(2011)\citenamefont {Wu},
  \citenamefont {Piper}, \citenamefont {Ediger}, \citenamefont {Brereton},
  \citenamefont {Schmidgall}, \citenamefont {Eastham}, \citenamefont {Hugues},
  \citenamefont {Hopkinson},\ and\ \citenamefont {Phillips}}]{wu2011pop}%
  \BibitemOpen
  \bibfield  {author} {
  \bibinfo {author} {\bibfnamefont {Y.}~\bibnamefont {Wu}}, 
  \bibinfo {author} {\bibfnamefont {I.~M.}\ \bibnamefont {Piper}},
  \bibinfo {author} {\bibfnamefont {M.}~\bibnamefont {Ediger}}, 
  \bibinfo {author} {\bibfnamefont {P.}~\bibnamefont {Brereton}}, 
  \bibinfo {author} {\bibfnamefont {E.~R.}\ \bibnamefont {Schmidgall}},
  \bibinfo {author} {\bibfnamefont {P.~R.}\ \bibnamefont {Eastham}},
  \bibinfo {author} {\bibfnamefont {M.}~\bibnamefont {Hugues}}, 
  \bibinfo {author} {\bibfnamefont {M.}~\bibnamefont {Hopkinson}}, \ and\
  \bibinfo {author} {\bibfnamefont {R.~T.}\ \bibnamefont {Phillips}},\ }\bibfield  {title} {\enquote {\bibinfo
  {title} {Population inversion in a single $\mathrm{InGaAs}$ quantum dot using the method of adiabatic rapid passage},}\ }\href@noop {} {\bibfield  {journal} {\bibinfo
   {journal} {Phys. Rev. Lett.}\ }\textbf {\bibinfo {volume} {106}},\ \bibinfo
  {pages} {067401} (\bibinfo {year} {2011})}\BibitemShut {NoStop}% 
\bibitem [{\citenamefont {Simon}\ \emph {et~al.}(2011)\citenamefont {Simon},
  \citenamefont {Belhadj}, \citenamefont {Chatel}, \citenamefont {Amand},
  \citenamefont {Renucci}, \citenamefont {Lema{\^i}tre}, \citenamefont {Krebs},
  \citenamefont {Dalgarno}, \citenamefont {Warburton}, \citenamefont {Marie},\
  and\ \citenamefont {Urbaszek}}]{simon2011rob}%
  \BibitemOpen
  \bibfield  {author} {
  \bibinfo {author} {\bibfnamefont {C.~M.}\ \bibnamefont {Simon}}, 
  \bibinfo {author} {\bibfnamefont {T.}~\bibnamefont {Belhadj}},
  \bibinfo {author} {\bibfnamefont {B.}~\bibnamefont {Chatel}}, 
  \bibinfo {author} {\bibfnamefont {T.}~\bibnamefont {Amand}}, 
  \bibinfo {author} {\bibfnamefont {P.}~\bibnamefont {Renucci}}, 
  \bibinfo {author} {\bibfnamefont {A.}~\bibnamefont {Lema{\^i}tre}},
  \bibinfo {author} {\bibfnamefont {O.}~\bibnamefont {Krebs}}, 
  \bibinfo {author} {\bibfnamefont {P.~A.}\ \bibnamefont {Dalgarno}},
  \bibinfo {author} {\bibfnamefont {R.~J.}\ \bibnamefont {Warburton}},
  \bibinfo {author} {\bibfnamefont {X.}~\bibnamefont {Marie}}, \ and\
  \bibinfo {author} {\bibfnamefont {B.}~\bibnamefont {Urbaszek}},\ }\bibfield  {title} {\enquote {\bibinfo {title} {Robust quantum
  dot exciton generation via adiabatic passage with frequency-swept optical
 pulses},}\ }\href@noop {} {\bibfield  {journal} {\bibinfo  {journal} {Phys. Rev. Lett.}\ }\textbf {\bibinfo {volume} {106}},\ \bibinfo {pages} {166801} (\bibinfo {year} {2011})}\BibitemShut {NoStop}%
\bibitem [{\citenamefont {Mathew}\ \emph {et~al.}(2014)\citenamefont {Mathew},
  \citenamefont {Dilcher}, \citenamefont {Gamouras}, \citenamefont
  {Ramachandran}, \citenamefont {Yang}, \citenamefont {Freisem}, \citenamefont
  {Deppe},\ and\ \citenamefont {Hall}}]{mathew2014sub}%
  \BibitemOpen
  \bibfield  {author} {
  \bibinfo {author} {\bibfnamefont {R.}~\bibnamefont {Mathew}}, 
  \bibinfo {author} {\bibfnamefont {E.}~\bibnamefont {Dilcher}},
  \bibinfo {author} {\bibfnamefont {A.}~\bibnamefont {Gamouras}}, 
  \bibinfo {author} {\bibfnamefont {A.}~\bibnamefont {Ramachandran}},
  \bibinfo {author} {\bibfnamefont {H.~Y.~S.}\ \bibnamefont {Yang}},
  \bibinfo {author} {\bibfnamefont {S.}~\bibnamefont {Freisem}}, 
  \bibinfo {author} {\bibfnamefont {D.}~\bibnamefont {Deppe}}, \ and\
  \bibinfo {author} {\bibfnamefont {K.~C.}\ \bibnamefont {Hall}},\ }\bibfield  {title} {\enquote {\bibinfo {title}
  {Subpicosecond adiabatic rapid passage on a single semiconductor quantum dot: Phonon-mediated dephasing in the strong-driving regime},}\ }\href@noop {}
  {\bibfield  {journal} {\bibinfo  {journal} {Phys. Rev. B}\ }\textbf {\bibinfo {volume} {90}},\ \bibinfo {pages} {035316} (\bibinfo {year}
  {2014})}\BibitemShut {NoStop}%
\bibitem [{\citenamefont {Eastham}\ \emph {et~al.}(2013)\citenamefont
  {Eastham}, \citenamefont {Spracklen}, \ and\
  \citenamefont {Keeling}}]{eastham2013lin}%
  \BibitemOpen   
\bibfield  {journal} {  }\bibfield  {author} {
  \bibinfo {author} {\bibfnamefont {P.~R.}\ \bibnamefont {Eastham}}, 
  \bibinfo {author} {\bibfnamefont {A.~O.}\ \bibnamefont {Spracklen}}, 
  \bibinfo {author} {\bibfnamefont {J.}~\bibnamefont {Keeling}},\ }\bibfield
  {title} {\enquote {\bibinfo {title} {Lindblad theory of dynamical decoherence of quantum-dot excitons},}\
  }\href@noop {} {\bibfield  {journal} {\bibinfo  {journal} {Phys. Rev. B}\
  }\textbf {\bibinfo {volume} {87}},\ \bibinfo {pages} {195306} (\bibinfo
  {year} {2013})}\BibitemShut {NoStop}%  
\bibitem [{\citenamefont {Wei}\ \emph {et~al.}(2014)\citenamefont {Wei},
  \citenamefont {He}, \citenamefont {Chen}, \citenamefont {Hu}, \citenamefont
  {He}, \citenamefont {Wu}, \citenamefont {Schneider}, \citenamefont {Kamp},
  \citenamefont {H{\"o}fling}, \citenamefont {Lu},\ and\ \citenamefont
  {Pan}}]{wei2014det}%
  \BibitemOpen
  \bibfield  {author} {
  \bibinfo {author} {\bibfnamefont {Y.-J.}\ \bibnamefont {Wei}}, 
  \bibinfo {author} {\bibfnamefont {Y.-M.}\ \bibnamefont {He}},
  \bibinfo {author} {\bibfnamefont {M.-C.}\ \bibnamefont {Chen}}, 
  \bibinfo {author} {\bibfnamefont {Y.-N.}\ \bibnamefont {Hu}}, 
  \bibinfo {author} {\bibfnamefont {Y.}~\bibnamefont {He}}, 
  \bibinfo {author} {\bibfnamefont {D.}~\bibnamefont {Wu}}, 
  \bibinfo {author} {\bibfnamefont {C.}~\bibnamefont {Schneider}}, 
  \bibinfo {author} {\bibfnamefont {M.}~\bibnamefont {Kamp}},
  \bibinfo {author} {\bibfnamefont {S.}~\bibnamefont {H{\"o}fling}},
  \bibinfo {author} {\bibfnamefont {C.-Y.}\ \bibnamefont {Lu}}, \ and\
  \bibinfo {author} {\bibfnamefont {J.-W.}\ \bibnamefont {Pan}},\ }\bibfield  {title} {\enquote
  {\bibinfo {title} {Deterministic and robust generation of single photons from a single quantum dot with 99.5\% indistinguishability using adiabatic rapid passage},}\ }\href@noop {} {\bibfield  {journal} {\bibinfo  {journal} {Nano
  Lett.}\ }\textbf {\bibinfo {volume} {14}},\ \bibinfo {pages} {6515}
  (\bibinfo {year} {2014})}\BibitemShut {NoStop}%
\end{thebibliography}
\end{document}